# Fisher information for quasi-one-dimensional hydrogen atom

Aparna Saha[1], Benoy Talukdar[1] and Supriya Chatterjee[2]

[1] Department of Physics, Visva-Bharati University, Santiniketan 731235, India

[2] Department of Physics, Bidhannagar College, EB-2, Sector-1, Salt Lake, Kolkata 700064, India

**Abstract**. The coordinate-space wave function $\psi(x)$ of quasi-one-dimensional atoms is defined in the $x \geq 0$ region only. This poses a typical problem to write a physically acceptable momentum-space wave function $\varphi(p)$ from the Fourier transform of $\psi(x)$. We resolve the problem with special attention to the behavior of real and imaginary parts of the complex-valued function $\varphi(p)$ as a function of $p$ and confirm that $\varphi_i(p)$ (the imaginary part of $\varphi(p)$) represents the correct momentum-space wave function. We make use of the results for $\psi(x)$ and $\varphi_i(p)$ to express the position- and momentum-space Fisher information in terms of the principal quantum number and energy eigen value of the system and provide some useful checks on the result presented with particular attention on the information theoretic uncertainty relation.

## 1. Introduction

Traditionally, information theory deals with quantification and storage of information to use them in communication systems. However, information theoretical studies play a role in many other applicative contexts. In the recent past, Frieden [1] established a direct connection between the information theory and basic principles of physics by deriving a mathematical procedure, currently known as 'the law of extreme physical information'. But curiously enough, the relation between information theory and quantum mechanics was noted long ago [2]. Particularly, application of information theory to quantum systems could explain basic features of many microscopic processes including the so-called quantum computation [3]. It is often believed that the celebrated work of Shannon [4] gave birth to what we currently call the 'information theory'. However, long before the publication of Shannon's work, Fisher [5] introduced a measure of information which goes by the name Fisher information. Both information measures depend on probability density corresponding to changes in some observable. The information entropy of Shannon is very little sensitive over a small-sized region. As opposed to this, Fisher information can detect local changes in the density distribution and thus provide better description of the system in an information theoretic way.

The connection between Fisher information and quantum theory can be established by using many elegant formal approaches like the de-quantization procedure [6]. Here the quantum kinetic energy is written as a classical term plus a purely quantum term [7]. The latter term which arises due to quantum fluctuation is identical to the Fisher information while the classical term plays a role analogous to that of the Shannon entropy. The object of the present work is to study the properties of position- and momentum-space Fisher information, $I_\rho$ and $I_\gamma$, for the quasi-one-dimensional (Q1D) hydrogen atom by using an approach which is physically transparent and mathematically rigorous. The classical phase-space structure of this simplified atomic model has been found to closely mimic the three-dimensional systems for initial conditions representing elongated Stark states [8]. On the other hand, the Q1D hydrogen atom provides a very useful basis to derive analytical approaches to study the dynamics of Rydberg wave packets with emphasis on their revival, super-revival and phase-space localization [9]. In the recent past, it has been found that wave packet revivals and fractional revivals can also be analyzed by using a measure of non-classicality based on the Fisher information [10]. Keeping these in view we shall first derive an analytical model to construct expressions for $I_\rho$ and $I_\gamma$ in terms of energy eigen values of the atom and then examine the uncertainty character of the Fisher information product $I_\rho I_\gamma$. In this context we note that, save the work of Romera et al [11], Fisher information of physical systems has hardly been examined from analytical standpoint.

There exist experimental techniques [12] which allow production of wave packets moving along elliptical orbits of arbitrary eccentricity ranging from a circle to a line. Understandably, the ellipse with the maximum eccentricity represents a line on the one side of the atomic core and leads to the formation of the so-called Q1D hydrogen atom. In Hartree units the eigen function of the Hamiltonian for this system is given by [13]

$$\psi(x) = \frac{2x}{\sqrt{n^3}} e^{-\frac{x}{n}} {}_1F_1\left(-n+1; 2; \frac{2x}{n}\right), \quad x \geq 0 \tag{1}$$

with the energy eigen value written as

$$E_n = -\frac{1}{2n^2}. \tag{2}$$

Here the principal quantum number $n = 1, 2, 3, etc$ and ${}_1F_1(.)$ stands for the confluent hypergeometric function. From (1) it is evident that $\psi(x)$ corresponds to the radial Schrödinger wave function of the three-dimensional hydrogen atom for angular momentum $l = 0$. In sec. 2 we shall see that because of the restriction on the domain of $x$ in (1) there appear certain typical difficulties to express $\psi(x)$ in the momentum space. Fortunately, these difficulties can be circumvented by paying special attention to the zeros of the real and imaginary parts of the Fourier transform of $\psi(x)$ constructed by restricting $x$ to the semi-infinite interval rather than $-\infty \leq x \leq \infty$. We carry out the analysis with sufficient care and thus provide an expression for the correct momentum-space wave function $\varphi(p)$ corresponding to $\psi(x)$ in (1). We then make use of these wave functions in sec.3 to construct exact analytical expressions for position- and momentum-space Fisher information, and subsequently examine the nature of the product $I_\rho I_\gamma$ as a function of the principal quantum number $n$. Finally, in sec.4 we summarize our outlook on the present work and make some concluding remarks.

2. **Momentum-space wave function**

For a true one-dimensional system $(-\infty \leq x \leq \infty)$ the momentum- and position-space wave functions are related by the Fourier transform

$$\varphi(p) = \frac{1}{\sqrt{2\pi}} \int_{-\infty}^{\infty} e^{-ipx} \psi(x) dx, \quad -\infty \leq p \leq \infty. \tag{3}$$

It may appear plausible to compute the momentum-space wave function for a Q1D atom $(0 \leq x \leq \infty)$ by replacing the infinite integral in (3) by an integral over the semi-infinite interval. This will permit use of the standard integral

$$\int_0^\infty e^{-\lambda x} x^\nu {}_1F_1(\alpha; \gamma; kx) dx = \frac{\Gamma(\nu+1)}{\lambda^{\nu+1}} {}_2F_1\left(\alpha, \nu+1; \gamma, \frac{k}{\lambda}\right), \quad \text{Re}\,\lambda > \text{Re}\,k > 0 \tag{4}$$

and the well known result ${}_2F_1(\alpha, \beta; \beta; x) = (1-x)^{-\alpha}$ for the Gaussian hypergeometric function to obtain the required momentum-space wave function in the form

$$\varphi(p) = (-1)^{n-1} \sqrt{\frac{2n}{\pi}} \frac{(1-inp)^{n-1}}{(1+inp)^{n+1}}. \tag{5}$$

The real and imaginary parts of the complex wave function in (5) can be written as

$$\phi_r(p) = (-1)^{n-1} \sqrt{\frac{2n}{\pi}} \frac{\cos(2n \arctan(np))}{1+n^2 p^2} \tag{6}$$

and

$$\varphi_i(p) = (-1)^n \sqrt{\frac{2n}{\pi}} \frac{\sin(2n \arctan(np))}{1+n^2 p^2}. \tag{7}$$

The coordinate-space wave function (1) vanishes at $x = 0$ and has additional $(n-1)$ zeros in the semi-infinite interval $0 \leq x < \infty$. Thus one would expect that a well-behaved momentum-space wave function will also exhibit a similar behavior as a function of $p$. We can use this criterion to choose the physically admissible momentum-space wave function. In order that we proceed by comparing the plots of $\varphi_r(p)$ and $\varphi_i(p)$ as a function of $p$ with the corresponding plot of $\psi(x)$. In Fig.1 we display the plot of $\psi(x)$ as a function of $x$ for $n = 4$ with the corresponding plots of $\varphi_r(p)$ and $\varphi_i(p)$ presented in Fig.2.

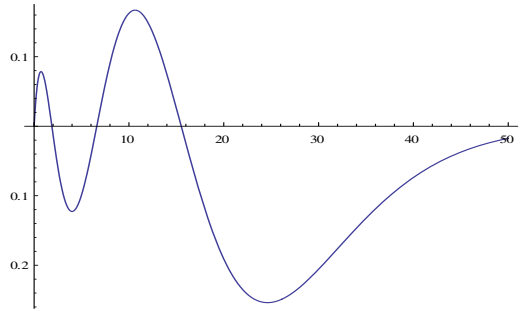

Fig.1 The coordinate-space wave function $\psi(x)$ as a function of $x$ for $n = 4$.

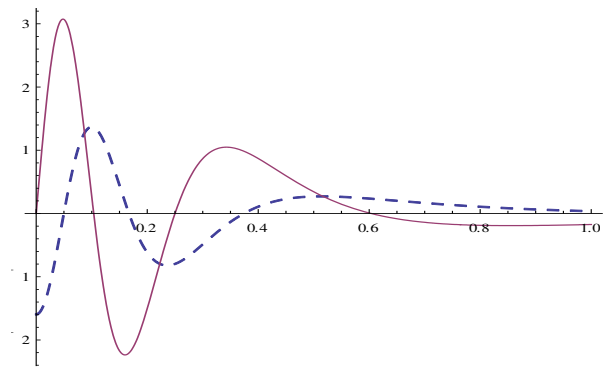

Fig.2 Real [$\varphi_r(p)$ dotted curve] and imaginary [$\varphi_i(p)$ solid curve] parts of the complex momentum-space wave function (5) as a function of $p$ for $n = 4$.

Looking at the curve in Fig.1 we see that the position-space wave function $\psi(x)$ is zero at $x = 0$ and, as expected, has three other additional zeros. On the other hand, the curve for the real part of the momentum-space wave function, $\varphi_r(p)$, in Fig.2 is non-zero (negative) at $p = 0$. It may appear that $\varphi_r(p)$ has also three nodes. But the innermost node arises from the unphysical behavior of $\varphi_r(p)$ at the origin and we should not treat it on equal footing with the outer ones. We can make this point more explicit with special attention to the curves in Fig.3 which portrays $\varphi_i(p)$ and $\varphi_r(p)$ for $n = 1$ as a function of $p$. For $n = 1$ the wave function should not have any node. As expected the imaginary part of the complex wave function, namely, $\varphi_i(p)$ in Fig.3 is nodeless but the real part $\varphi_r(p)$ has a physically unwanted node. We have verified that only $\varphi_i(p)$ exhibits the physically acceptable number of nodes for all values of $n$ and represents the true momentum-space wave function for the Q1D hydrogen atom.

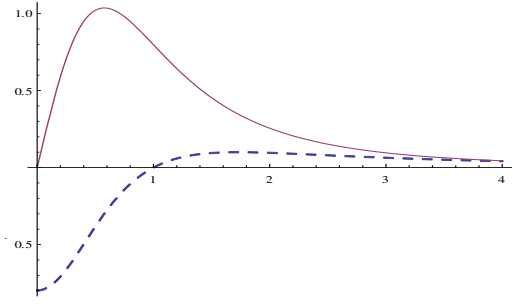

Fig.3 Real [$\varphi_r(p)$ dotted curve] and imaginary [$\varphi_i(p)$ solid curve] parts of the complex momentum-space wave function (5) as a function of $p$ for $n=1$.

We shall, therefore, make use of the position-space wave functions in (1) and properly normalized momentum-space wave function from (7) to express the corresponding Fisher information in terms of energy eigen values of one-dimensional hydrogen atom. In Appendix A we show that the expression for $\varphi_i(p)$ can also be obtained from the momentum-space hydrogenic wave function of Podolosky and Pauling [14] in the limit of zero angular momentum. This provides an additional support in favor of our claim that $\varphi_i(p)$ is the correct momentum-space wave function of the Q1D hydrogen atom.

3. **Fisher information**

The classical Fisher information for translations of a one-dimensional observable $X$ with corresponding probability density $\rho(x)$ is given by [15]

$$I_\rho = \int \frac{\rho'^2(x)}{\rho(x)} dx, \quad \rho'(x) = \frac{d\rho(x)}{dx}. \tag{8}$$

In (8) the limits of integration should be chosen to cover the infinite interval such that $-\infty \leq x \leq \infty$. The information measure (1) has a local character due to the presence of the gradient $(d/dx)$ operator and is sensitive to fluctuations in infinitesimally-neighboring $(dx)$ values of $\rho(x)$. The sensitivity is particularly strong because it depends on the square of the fluctuation. The primary application of Fisher information in classical estimation theory gives the lower bound

$$VarX \geq I_\rho^{-1} \tag{9}$$

for the variance of $X$. Relation (9) is known as the Cramer-Rao inequality [16]. In close analogy with position-space Fisher information one can introduce the momentum-space Fisher information written as

$$I_\gamma = \int \frac{\gamma'^2(p)}{\gamma(p)} dp, \quad -\infty \leq p \leq \infty, \tag{10}$$

where $\gamma(p)$ represents the probability density corresponding to translations of a one-dimensional observable $P$ in the momentum space.

Fisher information can be physically realized as a measure of disorder or smoothness of the probability density $\rho(x)$ and uncertainty of the associated random variable $X$. Frieden [17] envisaged detailed studies to investigate the disorder aspect. On the other hand, the uncertainty properties are clearly delineated by the Stam inequalities [18]. In close analogy with the stronger version of the Heisenberg uncertainty relation for Shannon's information entropy [19], the product $I_\rho I_\gamma$ has been conjectured to exhibit a non-trivial lower bound which for one-dimensional systems reads

$$I_\rho I_\gamma \geq 4. \tag{11}$$

The inequality in (11) is known as the Fisher information based uncertainty relation [20]. There are many quantum systems for which relation (11) is valid. There also exists a large number of counter examples which appear to demonstrate that it is not always possible to write a universal uncertainty relation as a lower bound to the product $I_\rho I_\gamma$ [21].

To calculate the results for $I_\rho$ and $I_\gamma$ for the present study we shall replace the infinite integrals in (8) and (10) by semi-infinite ones, and then make use of the position- and momentum-space probability densities $\rho(x) = \psi^2(x)$ and $\gamma(p) = \varphi_i^2(p)$ to compute results for Fisher information. The expression for $I_\rho$ is characterized by the following integrals

$$i_1 = \int_0^\infty x^m e^{-\frac{2x}{n}} {}_1F_1^2\left(1-n;2;\frac{2x}{n}\right), \quad m = 0,1,3, \tag{12}$$

$$i_2 = \int_0^\infty x^m e^{-\frac{2x}{n}} {}_1F_1\left(1-n;2;\frac{2x}{n}\right) {}_1F_1\left(2-n;3;\frac{2x}{n}\right), \quad m = 1,2 \tag{13}$$

and 
$$i_3 = \int_0^\infty x^2 e^{-\frac{2x}{n}} {}_1F_1^2\left(2-n;3;\frac{2x}{n}\right). \tag{14}$$

The integrands in (12) and (14) involve squares of the confluent hypergeometric function ${}_1F_1(.)$. As a result $i_1$ and $i_2$ can be evaluated directly by making use of [22]

$$\int_0^\infty e^{-\lambda z} z^{\nu-1} {}_1F_1^2(-n;\gamma;kz) dz = \frac{\Gamma(\nu) n!}{k^\nu \gamma(\gamma+1).......(\lambda+n-1)} S \tag{15}$$

with 
$$S = 1 + \frac{n)(n-\nu-1)(n-\nu)}{1^2 \gamma} + ............ . \tag{16}$$

The integral $i_2$ can also be evaluated with the help of (15) by writing the former in a suitable form given by

$$i_2 = \frac{n}{2(n-1)} Limit_{n\to 0} \frac{d}{d\varepsilon} \int_0^\infty x^{m-1} e^{-\frac{2x}{n}} {}_1F_1^2\left(1-n;2;\frac{2\varepsilon x}{n}\right) dx . \tag{17}$$

In writing (17) we have made use of the formula

$$_1F_1(a+1;c+1;z) = \frac{c}{a}\frac{d}{dz}\,_1F_1(a;c;z)\ .\tag{18}$$

Using the values of $i_1, i_2$ and $i_3$ we obtain a very simple expression for the position-space information written as

$$I_\rho = 8|E_n|\ .\tag{19}$$

The momentum-space probability density as obtained from the normalized wave function corresponding to that in (7) is given by

$$\gamma(p) = \frac{8n}{\pi}\frac{\sin^2\theta}{(1+n^2p^2)^2}\tag{20}$$

with 
$$\theta = 2n\arctan(np)\ .\tag{21}$$

We now restrict the variation of $p$ in the semi-infinite interval and make use of (10) and (20) to write

$$I_\gamma = \frac{128 n^5}{\pi}\int_0^\infty \frac{(\cos\theta - p\sin\theta)^2}{(1+n^2p^2)^4}dp\ .\tag{22}$$

Carrying out the elementary integrals in (22) we finally obtain

$$I_\gamma = \frac{5n^2+1}{|E_n|}\ .\tag{23}$$

For the quasi-one-dimensional hydrogen atom (8) and (10) can also be written as

$$I_\rho = 4\int_0^\infty \left|\frac{d}{dx}\rho^{\frac{1}{2}}(x)\right|^2 dx\ \ \text{and}\ \ I_\gamma = 4\int_0^\infty \left|\frac{d}{dp}\gamma^{\frac{1}{2}}(p)\right|^2 dp\ .\tag{24}$$

Replacing the probability densities in terms of wave functions we can recast (24) in the form

$$I_\rho = 4\int_0^\infty |\psi'(x)|^2 dx\ \ \text{and}\ \ I_\gamma = 4\int_0^\infty |\varphi'(p)|^2 dp\ .\tag{25}$$

Looking at (25) it appears that $I_\rho$ and $I_\gamma$ written in terms of derivatives of the position- and momentum-space wave functions will provide some calculational simplicity for the problem. One can easily verify that this is, however, not true.

The Fisher information and the associated uncertainty relation of single-particle systems with central potentials have been studied in both three and higher spatial dimensions [11,23]. We note that the results found for the three-dimensional case follow from the corresponding D-dimensional results for D=3. Thus we believe that, because of the following, a useful check on the results in (19) and (23) for the Q1D atom will consist in confronting them with the appropriate results obtained in the three-dimensional case.

The coordinate-space wave function (1) is just the hydrogenic radial wave function for the angular momentum $l=0$. Further, we have shown that the momentum-space wave function (7) is equal to the zero angular momentum wave function of Podolosky and Pauling [14]. Thus one would expect that the results in (19) and (23) will also follow from the corresponding hydrogenic results for $l=0$. From the expressions presented in ref. (11) by Romera et al we see this is indeed the case.

The uncertainty relation (11) involves Fisher information in two complementary spaces. In addition to single-particle systems in the central potential, this inequality has also been proved also for general mono-dimensional systems with even wave functions [24]. For Q1D hydrogen atom the product $I_\rho I_\gamma$ given by

$$I_\rho I_\gamma = 8(5n^2 + 1) \tag{26}$$

depends quadratically on the square of the principal quantum number. The lower bound obtained by putting $n=0$ is twelve times the lower bound predicted by (11) and, as expected, corresponds to the result in ref. 11 for the angular momentum $l=0$.

4. **Conclusion**

In this paper we presented a rigorous derivation of the results for position- and momentum-space Fisher information, $I_\rho$ and $I_\gamma$, for the Q1D hydrogen atom and subsequently examined the properties of the uncertainty relation involving them. The coordinate-space wave function $\psi(x)$ of this atom plays a crucial role in studying the revival, super-revival and fractional revival of Rydberg wave packets {13}. It appears the momentum-space wave function has not yet been used in any applicative context. Since electronic motion in Q1D atom is restricted to $x \geq 0$ region, one expects to construct the corresponding momentum-space wave function $\varphi(p)$ by restricting to the semi-infinite interval $0 \leq x \leq \infty$ while taking the Fourier transform of $\psi(x)$. This leads to an awkward physical constraint for the real part of $\varphi(p)$ such that $\varphi_i(p)$ represents the correct wave function for future application. We reconfirmed this fact by deriving the expression for $\varphi_i(p)$ from the momentum-space wave function of Podolosky and Pauling [14] for angular momentum $l=0$. In this context we note that the result for $\varphi_i(p)$ is given in terms of elementary transcendental functions only while that of ref. 14 as recently used by us [27] is characterized by Chebyshev polynomial. The wave functions $\psi(x)$ and $\varphi_i(p)$ (which is essentially the sine transform of $\psi(x)$) are then used to write the position- and momentum-space Fisher information in simple closed form which provided a basis for straightforward realization for the information-based uncertainty relation of the system. The treatment presented by us is expected to resolve any confusion regarding the choice of momentum-space wave function for the Q1D hydrogen atom.

**Acknowledgement**

The authors would like to thank Pranab Sarkar and Madan Mohan Panja for some useful discussion.

**Appendix A : Derivation of (7) from the momentums-pace hydrogenic wave function of ref. 14**

For $l=0$ the momentum-space wave function of Podolosky and Pauling [10] is given by

$$\varphi_{PP}(p) = \frac{2^{\frac{5}{2}} n^{\frac{3}{2}}}{\sqrt{\pi}} \frac{p}{(1+n^2 p^2)^2} U_{n-1}\left(\frac{1-n^2 p^2}{1+n^2 p^2}\right). \quad (A1)$$

Here $U_m(.)$ stands for the Chebyshev polynomial of the second kind. To obtain (7) from (A1) we begin by noting the relation between $U_m(\xi)$ and Gaussian hypergeometric function $_2F_1(a,b;c;\zeta)$ written as [25]

$$U_m(\xi) = (m+1)\,_2F_1\left(-m, m+2; \frac{3}{2}; \frac{1-\xi}{2}\right). \quad (A2)$$

For the choice $m = n-1$ and $\xi = (1-n^2 p^2)/(1+n^2 p^2)$, (A2) reduces to

$$U_{n-1}\left(\frac{1-n^2 p^2}{1+n^2 p^2}\right) = n\,_2F_1\left(-n+1, n+1; \frac{3}{2}; \frac{n^2 p^2}{1+n^2 p^2}\right). \quad (A3)$$

The Gaussian hypergeometrc function in the left side of (A3) can be written in terms of elementary transcendental functions by using [26]

$$_2F_1\left(1-\frac{r}{2}, 1+\frac{r}{2}; \frac{3}{2}, \eta^2\right) = \frac{\sin(r \arcsin \eta)}{r\eta\sqrt{1-\eta^2}}. \quad (A4)$$

Making use of $r = 2n$ and $\eta = np/\sqrt{1+n^2 p^2}$ in (A4) we get

$$_2F_1\left(-n+1, n+1; \frac{3}{2}; \frac{n^2 p^2}{1+n^2 p^2}\right) = \frac{(1+n^2 p^2)\sin(2n \arctan(np))}{n^2 p}. \quad (A5)$$

From (A3) and (A5) we can write

$$U_{n-1}\left(\frac{1-n^2 p^2}{1+n^2 p^2}\right) = \frac{(1+n^2 p^2)\sin(2n \arctan(np))}{np}. \quad (A6)$$

We can now combine (A1) and (A7) to obtain the normalized wave function associated with (7).

**References**

[1] B. R. Frieden, *Science from Fisher information*, Cambridge University Press, Cambridge, 2004.

[2] S. B. Sears, R. G. Parr and U. Dinir, Israel J. Chem. **19**, 165 (1980).

[3] M. A. Nielsen and I. L. Chuang, *Quantum Computation and Quantum Information*, Cambridge University Press, Cambridge, 2000.

[4] C. E. Shannon, Bell Syst. Tech. J. **27**, 379 (1948).

[5] R. A. Fisher, Cam. Phil. Soc**. 22**, 70000 (1925).

[6] R. A.Mosna, I. P. Hamilton and L. Delle Site, J. Phys. A : Math. Theor. **38**, 3869 (2005); J. Math. Phys. : Math. Theor**. 39**, L229 (2006).

[7] I. P. Hamilton and R. A. Mosna, arXiv : 0905.3539v1 [quant-ph] 21 May2009.

[8] M. T. Frey, F. B. Dunning, C. O. Reihold and J. Burgdorfer, Phys. Rev. A **59**, 1434 (1999).

[9]F. B. Dunning, J. J. Mestayer, C. O. Reinhold, S. Yoshida and J. Burgdorfer, J. Phys. B : At. Mol. Opt. Phys. **42** 022001 (2009); Supriya Chatterjee, Aparna Saha and B Talukdar, Eur. Phys. J. **D67**, 240 (2013).

[10] E. Romera and F. de los Santos, Fisher information, arXiv : 1409.5599v1 [quant-ph] 19 Sep 2014.

[11] E. Romera, P. Sanchez-Moreno and J. S. Dehesa, Chem. Phys. Lett. **414**, 468 (2005).

[12] J. Bromage and C. R. Stroud Jr. ,Phys. Rev. Lett. **83**, 4963(1970).

[13] I. Bersons and R. Veilande, Phys. Rev. A **69**, 043408 (2004); A. Saha, S. Chatterjee and B. Talukdar, Phys. Scr **81**, 055302 (2010); S. Chatterjee, A. Saha and B. Talukdar, Acta Phys. Polo**. A120**, 483 (2011).

[14] B. Podolsky and L Pauling 1939 Phys. Rev. **34,** 109 (1939).

[15] R. A. Fisher. Proc. Camb. Phil. Soc. 22, 700 (1925).

[16] D. R. Cox and D. V. Hinkely, *Theoretical Statistics* ( Chapman and Hall, London 1974).

[17] B. R. Frieden, Phys Rev. A **41**, 4265 (1990).

[18] A. Stam, Inform. Control **2**, 105 (1959).

[19] I.Bialynicki-Birula and J.Myceilski, Comm. Math. Phys. **44**, 129 (1975).

[20] J. S. Dehesa, R. Gonzalez-Ferez and P. Sanchez-Moreno, J. Phys. A : Math. Theor. **40**, 1845 (2007).

[21] V. Aguiar and I.Guedes, Phys. Scr**. 98**, 045207 (2015) ; A. Plastino, G. Bellomo and A. R. Plastino, Adv. Math. Phys.**1**, 215 (2015).

[22] L. D. Landau and E. M. Lifshitz, *S*, 2[nd] Edn, Ed. J. B. Sykes and J. S. Bell (Pergamon, New York 1965).

[23] E. Romera, P. Sanchez-Moreno and J. S. Dehesa, J. Math. Phys. 47, 103504 (2006).

[24] J. S. Dehesa, A. Martinez-Finkelshtien and V. N. Sorokins, , Mol. Phys.**10**4, 613 (2006).

[25] M. Abramowitz and I.Stegun *Handbook of Mathematical Physics* (Dover publication, New York,1972) pp 779.

[26] W. Magnus and F. Oberhettinger, *Formulas and Theorems for the Special Functions of Mathematical Physics* ( translated by John Wermer) (Chelsea Pub. Co., New York,1949) pp8.
[27] Aparna Saha, Benoy Talukdar and Supriya Chatterjee, Eur. J. Phys. 38, 025103 (2017).